\documentclass[runningheads]{llncs}

\usepackage{graphicx}
\usepackage[protrusion=true,spacing=true]{microtype}
\usepackage{tikz}
\usepackage{amsmath}
\usepackage{amsfonts}
\usepackage{mathtools}  
\usepackage{caption}
\usepackage{subcaption} 
\usepackage{booktabs}
\usepackage{paralist}
\usepackage{siunitx}
\usepackage[bookmarks=true]{hyperref}
\hypersetup{bookmarksdepth=2}
\usepackage{url}

\newcommand{\actionset}{{\mathcal{A}}}

\newcommand{\ami}{{\textsc{ami}}}
\newcommand{\pdt}{{\textsc{pdt}}}
\newcommand{\pdts}{{\textsc{pdt}s}}
\begin{document}
%
\title{Towards Quantum-Secure Authentication and Key Agreement via Abstract Multi-Agent Interaction}
\titlerunning{Secure Authentication and Key Agreement via Multi-Agent Interaction}

\author{
Ibrahim H. Ahmed
\and
Josiah P. Hanna
\and
Elliot Fosong
\and
Stefano V. Albrecht 
}
\authorrunning{I. H. Ahmed et al.}
\institute{
School of Informatics, University of Edinburgh, Edinburgh EH8 9AB, UK \\
\email{\{i.ahmed, josiah.hanna, e.fosong, s.albrecht\}@ed.ac.uk}
}
\maketitle

\begin{abstract}
Current methods for authentication and key agreement based on public-key cryptography are vulnerable to quantum computing. We propose a novel approach based on artificial intelligence research in which communicating parties are viewed as autonomous agents which interact repeatedly using their private decision models. Authentication and key agreement are decided based on the agents' observed behaviors during the interaction. The security of this approach rests upon the difficulty of modeling the decisions of interacting agents from limited observations, a problem which we conjecture is also hard for quantum computing. 
We release PyAMI, a prototype authentication and key agreement system based on the proposed method.
We empirically validate our method for authenticating legitimate users while detecting different types of adversarial attacks.
Finally, we show how reinforcement learning techniques can be used to train server models which effectively probe a client's decisions to achieve more sample-efficient authentication.

\end{abstract}


\section{Introduction}
Authentication and key agreement protocols are the foundation for secure communication over computer networks.
Most protocols in use today are based on public-key cryptographic methods such as Diffie-Hellman key exchange, the RSA cryptosystem, and elliptic curve cryptosystems \cite{bernstein2009post-quantum}.
These methods rely on the difficulty of certain number theoretic problems which can be solved efficiently using quantum computing \cite{shor1999polynomial-time}.
Thus, researchers are studying alternative mathematical problems believed to be safe against quantum computing \cite{bernstein2009post-quantum}.
Standards organizations such as the US National Institute of Standards and Technology~\cite{nist2016} are calling for new quantum-safe proposals for standardization.

We propose a novel formulation of authentication and key agreement inspired by research in artificial intelligence (AI) and machine learning. In the proposed method, communicating parties are viewed as {\it autonomous agents} which interact repeatedly using their private decision models. Authentication and key agreement are 
based solely on the agents recognizing each other from their observed behavior, and no private information is sent at any time during the process.
Our approach creates a bridge to AI research in two ways:

\textbf{Security} -- 
The method's security rests upon the difficulty of modeling an agent's decisions from limited observations about its behavior -- a long-standing problem in AI research known as \emph{opponent modeling} \cite{albrecht2018autonomous}.
We conjecture that the problem is as hard for quantum computing, since the problem is fundamentally one of missing information regarding the causality in an agent's decisions (details in Sec. \ref{sec:auth_multiagent_interaction}). There are no known quantum algorithms to solve opponent modeling; indeed, if such an algorithm was invented as an attack on our method, it could provide significant novel insights for AI research.

\textbf{Optimization} -- By formulating authentication as a multi-agent interaction process, we can employ concepts and algorithms for \emph{optimal decision-making} from reinforcement learning (\textsc{rl}) \cite{sutton1998reinforcement} to optimize the efficiency of the process. The idea is to enable communicating agents to be strategic about probing each other's reactions to maximize authentication accuracy and efficiency. We apply \textsc{rl} methods to our framework to optimize the agent models to reduce the number of interactions required to reach high-confidence authentication decisions.

In summary, our contributions are the following.
We introduce a protocol for secure authentication and key agreement based on recognizing an agent from limited observations of its actions.
We show empirically that our method obtains high accuracy in rejecting different categories of adversarial agents, while accepting legitimate agents with high confidence.
We release a prototype implementation of this protocol, called PyAMI, which allows remote machines to authenticate to one another and generate symmetric session keys.
Finally, we introduce an approach for optimizing security based on \textsc{rl} and show empirically that it leads to a significantly more efficient protocol in terms of the required number of client/server interactions than a default random probing server agent.

\section{Related Work}

\textbf{Post-quantum alternatives:} 
Among current post-quantum methods in the literature, those based on the fields of coding theory \cite{stern_syndrome_decoding}, lattice theory \cite{micciano_vadhan}, and multivariate quadratic polynomials \cite{sakumoto} provide existing entity identification schemes. Such schemes avoid quantum vulnerability by relying on problems for which there is no known quantum algorithm. The use of optimization and decision-theoretic principles, however, makes our approach fundamentally different to other lines of investigation in post-quantum security which rely primarily on the development of new cryptographic operators.

\textbf{Symmetric AKE:} Protocols for symmetric authenticated key-exchange (AKE) such as Kerberos \cite{kerberos_rfc4120} often rely on a third party to provide session keys. Their session key may also be generated independently of the long-term key (LTK). In our protocol, parties generate session keys without the aid of an extra entity, and derive it based on the LTK. With respect to authentication, protocols like \cite{avoine_2019_SAKE} often use a MAC tag based on the LTK, while our protocol uses a test of statistical similarity to determine whether a party possesses the expected LTK.

\textbf{Information-theoretic secrecy:} 
Information-theoretic protocols rely on security which can be achieved without any assumptions on an attacker's computational limits. 
Shannon's introduction of such protocols required a shared secret key between communicating parties over a noiseless channel \cite{shannon_theory}. 
Later protocols replaced this requirement of a shared key by introducing stochasticity \cite{Wyner_wiretap}. 
Our key agreement protocol is similar to Shannon's original setting, relying upon a shared secret in the form of the client's decision model, but it is instead used to generate the session key itself for symmetric encryption and decryption.

\textbf{Multi-agent modeling / interactive processes}: Agent-based modeling has been applied quite broadly in the field of security, such as for analyzing dynamics between parties in a computer network \cite{wagner_lippmann}. Our protocol is a novel application of multi-agent theory and optimization to cryptographic authentication. Game-theoretic approaches, particularly \textit{security games}, have also been proposed for cyber-defense scenarios between attacker and defenders \cite{manshaei_et_al_survey}. Our own work does not rely on equilibrium concepts which are difficult to scale \cite{daskalakis2009complexity} and based on normative rationality assumptions.

\section{Authentication via Multi-agent Interaction}
\label{sec:auth_multiagent_interaction}

This section details our proposed protocol, called \textbf{A}uthentication via \textbf{M}ulti-agent \textbf{I}nteraction (\ami; pronounced ``Am I?"). In the following, we use calligraphic letters (e.g., $\mathcal{X}$) to denote sets, lower case letters to denote elements of sets and functions, and upper case letters to denote random variables.
We use $\Delta(\mathcal{X})$ to denote the set of all probability distributions over elements of set $\mathcal{X}$.

We consider a setting in which a client seeks to authenticate to a server as a particular user, $u$.
The server must decide whether the client is the intended (\emph{legitimate}) user $u$ or an \emph{adversarial} client attempting to access the server as the intended user.

\begin{figure}[t]
    \begin{center}
        \scalebox{0.80}{
\ifx\du\undefined
  \newlength{\du}
\fi
\setlength{\du}{15\unitlength}
\begin{tikzpicture}
\pgftransformxscale{1.000000}
\pgftransformyscale{-1.000000}
\definecolor{dialinecolor}{rgb}{0.000000, 0.000000, 0.000000}
\pgfsetstrokecolor{dialinecolor}
\definecolor{dialinecolor}{rgb}{1.000000, 1.000000, 1.000000}
\pgfsetfillcolor{dialinecolor}
\definecolor{dialinecolor}{rgb}{1.000000, 1.000000, 1.000000}
\pgfsetfillcolor{dialinecolor}
\fill (33.000000\du,10.000000\du)--(33.000000\du,18.600000\du)--(38.000000\du,18.600000\du)--(38.000000\du,10.000000\du)--cycle;
\pgfsetlinewidth{0.100000\du}
\pgfsetdash{}{0pt}
\pgfsetdash{}{0pt}
\pgfsetmiterjoin
\definecolor{dialinecolor}{rgb}{0.000000, 0.000000, 0.000000}
\pgfsetstrokecolor{dialinecolor}
\draw (33.000000\du,10.000000\du)--(33.000000\du,18.600000\du)--(38.000000\du,18.600000\du)--(38.000000\du,10.000000\du)--cycle;
\definecolor{dialinecolor}{rgb}{0.000000, 0.000000, 0.000000}
\pgfsetstrokecolor{dialinecolor}
\node at (35.500000\du,14.20000\du){\parbox{4.3em}{\textbf{Client} \\ model: $\pi_c$}};
\definecolor{dialinecolor}{rgb}{1.000000, 1.000000, 1.000000}
\pgfsetfillcolor{dialinecolor}
\fill (50.000000\du,10.000000\du)--(50.000000\du,18.600000\du)--(55.000000\du,18.600000\du)--(55.000000\du,10.000000\du)--cycle;
\pgfsetlinewidth{0.100000\du}
\pgfsetdash{}{0pt}
\pgfsetdash{}{0pt}
\pgfsetmiterjoin
\definecolor{dialinecolor}{rgb}{0.000000, 0.000000, 0.000000}
\pgfsetstrokecolor{dialinecolor}
\draw (50.000000\du,10.000000\du)--(50.000000\du,18.600000\du)--(55.000000\du,18.600000\du)--(55.000000\du,10.000000\du)--cycle;
\definecolor{dialinecolor}{rgb}{0.000000, 0.000000, 0.000000}
\pgfsetstrokecolor{dialinecolor}
\node at (52.500000\du,14.20000\du){\parbox{4.5em}{\textbf{Server} \\ model: $\pi_s$}};
\pgfsetlinewidth{0.100000\du}
\pgfsetdash{}{0pt}
\pgfsetdash{}{0pt}
\pgfsetbuttcap
{
\definecolor{dialinecolor}{rgb}{0.000000, 0.000000, 0.000000}
\pgfsetfillcolor{dialinecolor}
\pgfsetarrowsstart{stealth}
\definecolor{dialinecolor}{rgb}{0.000000, 0.000000, 0.000000}
\pgfsetstrokecolor{dialinecolor}
\draw (50.000000\du,10.600000\du)--(38.000000\du,10.600000\du);
}
\definecolor{dialinecolor}{rgb}{0.000000, 0.000000, 0.000000}
\pgfsetstrokecolor{dialinecolor}
\node[anchor=west] at (42.600000\du,10.100000\du){User: $u$};
\pgfsetlinewidth{0.100000\du}
\pgfsetdash{}{0pt}
\pgfsetdash{}{0pt}
\pgfsetbuttcap
{
\definecolor{dialinecolor}{rgb}{0.000000, 0.000000, 0.000000}
\pgfsetfillcolor{dialinecolor}
\pgfsetarrowsstart{stealth}
\pgfsetarrowsend{stealth}
\definecolor{dialinecolor}{rgb}{0.000000, 0.000000, 0.000000}
\pgfsetstrokecolor{dialinecolor}
\draw (50.000000\du,13.000000\du)--(38.000000\du,13.000000\du);
}
\definecolor{dialinecolor}{rgb}{0.000000, 0.000000, 0.000000}
\pgfsetstrokecolor{dialinecolor}
\node[anchor=west] at (41.800000\du,12.400000\du){$t = 0$:   $\ A_c^0 \ \ A_s^0$};
\pgfsetlinewidth{0.100000\du}
\pgfsetdash{}{0pt}
\pgfsetdash{}{0pt}
\pgfsetbuttcap
{
\definecolor{dialinecolor}{rgb}{0.000000, 0.000000, 0.000000}
\pgfsetfillcolor{dialinecolor}
\pgfsetarrowsstart{stealth}
\pgfsetarrowsend{stealth}
\definecolor{dialinecolor}{rgb}{0.000000, 0.000000, 0.000000}
\pgfsetstrokecolor{dialinecolor}
\draw (50.000000\du,15.800000\du)--(38.000000\du,15.800000\du);
}
\definecolor{dialinecolor}{rgb}{0.000000, 0.000000, 0.000000}
\pgfsetstrokecolor{dialinecolor}
\node[anchor=west] at (41.800000\du,15.200000\du){$t = l$:   $\ A_c^l \ \ A_s^l$};
\definecolor{dialinecolor}{rgb}{0.000000, 0.000000, 0.000000}
\pgfsetstrokecolor{dialinecolor}
\node[anchor=west] at (43.800000\du,13.800000\du){$\vdots$};
\pgfsetlinewidth{0.100000\du}
\pgfsetdash{}{0pt}
\pgfsetdash{}{0pt}
\pgfsetbuttcap
{
\definecolor{dialinecolor}{rgb}{0.000000, 0.000000, 0.000000}
\pgfsetfillcolor{dialinecolor}
\pgfsetarrowsend{stealth}
\definecolor{dialinecolor}{rgb}{0.000000, 0.000000, 0.000000}
\pgfsetstrokecolor{dialinecolor}
\draw (50.000000\du,18.200000\du)--(38.000000\du,18.200000\du);
}
\definecolor{dialinecolor}{rgb}{0.000000, 0.000000, 0.000000}
\pgfsetstrokecolor{dialinecolor}
\node[anchor=west] at (39.500000\du,17.650000\du){Auth: yes iff. $(A_c^0,...,A_c^l) \sim \pi_u$};
\definecolor{dialinecolor}{rgb}{0.000000, 0.000000, 0.000000}
\pgfsetstrokecolor{dialinecolor}
\node at (35.400000\du,17.650000\du){\parbox{5em}{session key: \\ $\mathtt{key}(H_l,\pi_c)$}};
\definecolor{dialinecolor}{rgb}{0.000000, 0.000000, 0.000000}
\pgfsetstrokecolor{dialinecolor}
\node at (52.400000\du,17.650000\du){\parbox{5em}{session key: \\ $\mathtt{key}(H_l,\pi_u)$}};
\end{tikzpicture}}
    \end{center}
    \caption{Multi-agent Interaction Protocol}
    \label{fig:maap}
\end{figure}
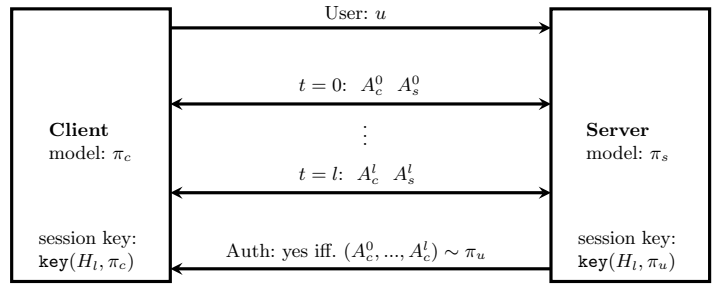

\textbf{Protocol:} When a client seeks to authenticate, the server initiates an interaction process which proceeds through time steps $t = 0,1,2,....,l$ (cf. Figure~\ref{fig:maap}).
At each time step $t$, the client and server independently choose actions $A_c^t$ and $A_s^t$, respectively, with values in a finite set of available actions, $\actionset \coloneqq \{1,...,n\}$.
The agents then send their chosen actions to each other.
The server associates a probabilistic decision model, $\pi_u$, with each legitimate user; the decision model is known only to the server agent and the legitimate user.
At the end of the interaction process, the server decides whether the interaction history $H_l \coloneqq (A_s^0, A_c^0, ..., A_s^l, A_c^l)$ was generated with a client using the model $\pi_u$ associated with the legitimate user.
If the server decides it has been interacting with this model, then it authenticates the client as user $u$; otherwise, it rejects the client agent.

We formalize agent decision models as functions mapping the past interaction history to a distribution over the available actions.
That is, the client chooses actions with the model $\pi_c: \mathcal{H}_t \rightarrow \Delta(\actionset)$ where $\mathcal{H}_t$ is the set of possible interaction histories up to time $t$. Similarly, the server agent chooses actions with a model $\pi_s$.
Various model representations could be used, including probabilistic decision trees, probabilistic finite state automata, and neural networks.
Jointly, the server and client agent define a distribution on $(l+1)$-step interaction histories, $p_{s,c}$.
If the client in the interaction process is a legitimate client then they use the decision model $\pi_u$ (i.e., $\pi_c = \pi_u$) that is a shared secret between the server and legitimate client.
To perform authentication, the server decides whether a given interaction history has been produced by $p_{s,u}$ or not.
To do so, we equip \ami\ with a test function, $\mathtt{test}: \mathcal{H}_l \rightarrow \{0, 1\}$ that returns $1$ if and only if an interaction $H \sim p_{s,u}$.
In Section \ref{sec:hypothesis} we describe how this function can be implemented with a frequentist hypothesis test.

\textbf{Key agreement:}
If the client is successfully authenticated, a secret session key can be computed as a function
$\mathtt{key}(H_l, \pi)$, where the client uses $\mathtt{key}(H_l, \pi_c)$ and the server uses $\mathtt{key}(H_l, \pi_u)$; if $\pi_c = \pi_u$, then the computed keys will be equal.
One possible implementation of the key function is by concatenating the probabilities $\pi_u(A_c^t | H_l), t=0,...l$ and pushing the resulting bit-string through a suitable hash function to produce a key with a desired length.

\textbf{Forward secrecy:}
\ami\ supports \emph{forward secrecy} \cite{gunther1990identity} to ensure that a compromised (e.g. stolen) legitimate user model cannot be used to compute past session keys. \ami\ transforms $\pi_u$ after each successful authentication process, such that the new model is a function $\pi_u' = \phi(\pi_u,H_l)$ of the old model, and such that $\phi$ is hard to invert. One possible transformation is to first reset a random seed to the value of the session key. Then, for each $\tau=0,...,l$, resample a new probability distribution for $\pi_u(H_{\tau})$. Since server and client use the same seed, they produce identical models $\pi_{u}'$ and $\pi_{c}'$. The session key cannot be recovered from a transformed model except by exhaustive search in the space of random seeds - with a sufficiently large key size, this is computationally infeasible
\cite{schneier_1st_edition}.

\textbf{Extension to more than two agents:}
\ami\ also supports mutual group authentication in which more than two parties authenticate to each other.
In this case, each agent $i = 1, ..., m$ has its own model $\pi_i$ which is a shared secret with all other legitimate agents.
The models are now defined over interaction histories which include the chosen actions of all agents at each time step, ($A_1^t, ..., A_m^t$).
Each agent authenticates each other agent using an authentication test, and the key function is similarly defined over all models, $\mathtt{key}(H_l, \pi_1, ..., \pi_m)$. In the remainder of this paper, we will focus on the basic setting in which a single client only authenticates to a single server.

\textbf{PyAMI Open-Source Framework:}
Towards further research on and adoption of \ami\ as a quantum-secure authentication protocol, we have developed an open-source Python application, PyAMI%
\footnote{PyAMI code and documentation: \texttt{\url{https://github.com/uoe-agents/PyAMI}}}%
. PyAMI consists of a multi-agent system where agents run on separate (virtual) machines, and communicate to authenticate over network sockets using TCP. 
During an interaction process, server and client machines transmit actions over a network to build the shared interaction history. After successful authentication, both parties compute identical session keys using the key agreement algorithm.

\section{Authentication via Hypothesis Testing}\label{sec:hypothesis}

To provide high-confidence authentication decisions, \ami\ uses the framework of frequentist hypothesis testing to decide whether a given interaction history was generated between the server and a legitimate client or an adversarial client.
For a given history, $h$, we first specify the null hypothesis ``$h$ was generated from $\pi_u$."
To decide on the correctness of this hypothesis, we compute a test statistic from the interaction history and determine whether the test statistic value is too extreme for the distribution of the test statistic under the null hypothesis.
More formally, letting $z: \mathcal{H} \rightarrow \mathbb{R}$ denote a test statistic function, a hypothesis test computes the $p$-value
\begin{equation}
p \coloneqq \Pr(|z(H)| \geq |z(h)|), \quad H \sim p_{s,u}.
\end{equation}

Intuitively, $p$ is the probability of observing a $z$ value at least as extreme as $z(h)$ if interacting with the legitimate client model.
The $p$-value is then compared to a pre-determined significance level, $\alpha$, to determine whether the interaction came from the legitimate client or not:
\begin{equation}
    \mathtt{test}(h) = 
    \begin{cases}
    \text{1 (authenticate)}  & \text{if $p$-value $\geq \alpha$ } \\
    \text{0 (reject)} & \text{if $p$-value $< \alpha$ }.
    \end{cases}
\end{equation}

We use a hypothesis test which was designed for non-stationary multi-agent interaction \cite{albrecht2015are}. 
Essentially, this test defines a flexible test statistic for multi-agent interaction, learns the distribution of this test statistic during an interaction (we use the score functions defined in \cite{albrecht2015are}), and computes $p$ from the learned distribution. Our only modification from the original algorithm is to fit the distribution of the test statistic with a normal distribution rather than a skew-normal distribution. This change allows us to compute p-values using the analytic normal CDF instead of the ratio-approximation proposed in \cite{albrecht2015are}, which led to more accurate results in our experiments.

An important aspect of the hypothesis testing approach is its interpretability.
The $p$-value has a well-defined semantics and the significance level $\alpha$ allows us to exactly control the false negative rate of the test. Under the null-hypothesis $\pi_c = \pi_u$, $p$ is uniformly distributed in $[0,1]$ and so a false negative occurs at exactly the rate $\alpha$.
If the legitimate client is incorrectly rejected, the client can retry the interaction process. The probability of $k$ successive false negatives is $\alpha^{k}$ which rapidly goes to zero.

\section{Protocol Security} 
\label{sec:protocol_security}

The problem of modeling the behavior of another agent from limited observations of its actions is widely studied in the AI research literature and known to be hard \cite{albrecht2018autonomous}.
The problem is fundamentally one of missing information regarding the causality in an agent's decisions, and this information can be difficult to extract from limited observations.
Even with a publicly known agent model structure -- which this paper assumes -- a complex model will involve large parameter spaces; inferring exact parameter values from a few observed authentications is infeasible.
The use of a quantum computer over a classical one will not aid in solving this specific type of problem, as it is more aligned with an \textit{information-theoretic} type of hardness rather than computational hardness \cite{Maurer1993TheRO}.

An information-theoretic key agreement protocol is considered (weakly) secure if: (1) the two parties' generated session keys agree with very high probability, (2) the key is nearly uniformly distributed, and (3) is nearly statistically independent of the information leaked to an intruder \cite{Maurer1993TheRO}. 
\ami\ is a symmetric key protocol and mandates that client and server generate identical session keys, fulfilling the first condition.

Regarding the second condition; in an experimental setting, \ami\ uses random instantiation so that the choice of user and server model is uniformly distributed over the space of possible models, which is significant as the session key is a function, $\mathtt{key}(H_l, \pi)$, of these models. 
Additionally, this key generation procedure includes a hash function as a final step -- we note that it is possible to also use a universal hashing mechanism here, similar to \cite{bennet_privacy} where universal hashing is applied so that possible outputs are equiprobable for an intruder. 

With respect to the third condition, \ami\ limits the publicly observable information by which an intruder may attempt to reconstruct $\pi_u$ and generate the correct session key. It does this in two ways -- first, it limits the length of the public interaction required for successful authentication (see optimization in Sec.~\ref{sec:empirical:probing}). Second, it implements a forward secrecy transform intended to limit all observations from a specific model $\pi_u$ to a single interaction session. The only way an intruder may obtain more than a single history from the same client model is in the unlikely event of a false negative, in which a legitimate client is incorrectly rejected (see Sec. \ref{sec:hypothesis}). 
We provide an empirical study of such a scenario in Section~\ref{sec:empirical:authentication} to demonstrate how \ami\ is robust against a maximum likelihood estimation (MLE) attack\footnote{Assuming a uniform prior distribution over possible models $\pi_u$, the best estimate of $\pi_u$ an attacker can formulate is the MLE; MLE is generally a preferred estimator among frequentist methods due to its statistical and asymptotic properties \cite{eliason_mle}.} even in the absence of the forward secrecy feature.

\section{Empirical Study: Authentication}\label{sec:empirical:authentication}

We now present an empirical study of the \ami\ protocol.
Our experiments are primarily designed to answer the following questions: 
\begin{inparaenum}[1)]
    \item Does \ami\ correctly accept a legitimate client?
    \item Does \ami\ correctly reject adversarial clients?
    \item How does the length of interaction histories affect \ami's accuracy?
    \item How robust is \ami\ to Maximum Likelihood Estimation attacks?
    \item How much time does PyAMI need to complete an interaction process?
\end{inparaenum}

\subsection{Authentication Empirical Set-up}
\label{auth_empirical_setup}

In our basic empirical setting, agents choose actions from $\actionset = \{1, ... , 10\}$.
The server model and legitimate client model are probabilistic decision trees (\pdts) --
decision trees in which each node has a probability distribution over actions.
The tree is traversed using the $k=5$ most recent actions of the other agent (i.e., the client tree is traversed with the server's actions).
We choose \pdts\ as they are computationally cheap to sample actions from and easy to randomly generate. 

For each experimental trial run, we randomly generate the server and true user decision model by setting each node in the \pdt\ to be a softmax distribution with logit values sampled uniformly in $[0,1]$ and temperature parameter $\tau$. 
The server decision model uses the value $\tau = 1.0$ for near-uniform random action selection; the client uses $\tau = 0.1$.
We find lower entropy in the client's action selection leads to better authentication accuracy with shorter interaction lengths. 
In each experimental trial, we generate interaction histories between the server and legitimate client and measure accuracy of the decisions made by \ami.
We also evaluate interactions between the server agent and adversarial agents. We formulate the following adversarial behaviors to create such interactions:

    \textbf{Random:}
    Generate a random adversarial \pdt\ with the same dimensions and temperature $\tau$ as the legitimate client \pdt.
    
    \textbf{Replay:}
     Replay client actions from observed interactions between the legitimate client and server to create adversarial ``replayed'' interaction histories. 
    
     \textbf{Maximum Likelihood Estimation (MLE-k):}
     Compute a maximum likelihood estimate of the legitimate client \pdt\ based on $k$ complete interaction histories with the legitimate client, assuming an identical \pdt\ structure. We set $k=100$ in these experiments.

To evaluate \ami\ we generate $1000$ interaction histories between the server and legitimate client, and $1000$ interaction histories between the server and each type of adversarial behavior for varying interaction history lengths.
We report authentication accuracy on each set of interaction histories as the percentage of interaction histories correctly identified as either legitimate or adversarial (Random, Replay, or MLE). For experimental rigor, we repeat this process over 100 different server and legitimate client models, and present the averaged results in Figure \ref{fig:length}.
For the hypothesis test we use a significance level of $\alpha=0.1$.

\subsection{Authentication Empirical Results}

\begin{figure}[t]
    \centering
    \includegraphics[scale=0.35, width=0.5\textwidth]{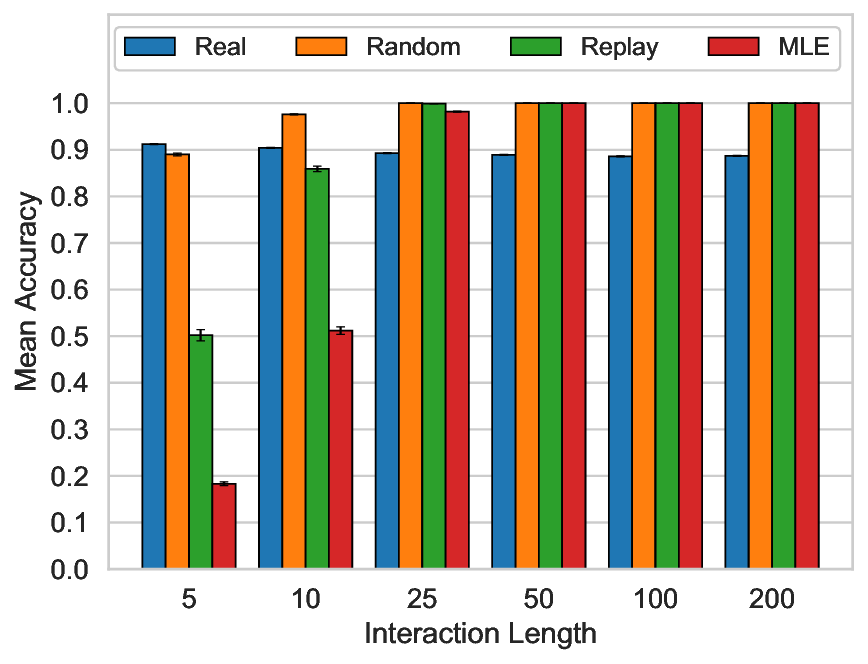}
    \caption{Authentication accuracy as a function of interaction length. For each considered interaction length we evaluate accuracy on every metric. For Real, Replay, and MLE metrics, results show accuracy on 1000 histories, averaged over 100 different server and legitimate client pairs. For the Random metric, results show accuracy on 1000 histories, averaged over 100 different server models.}
    \label{fig:length}
\end{figure}

Figure \ref{fig:length} shows the empirical accuracy of \ami\ with the legitimate client and against each type of adversary for a varying number of interaction history lengths.
As expected, the accuracy for the legitimate client model is unaffected by the interaction history length and always remains around $0.9$, due to our chosen significance threshold. 
For shorter history lengths ($l<50$), MLE is the strongest attack tested.
Once interaction histories are sufficiently long ($l \geq 50$), however, accuracy is perfect against adversarial clients and nothing is gained by further increasing the interaction history length.
We emphasize that the MLE adversary cannot successfully authenticate \emph{even after observing 100 interaction histories from the legitimate client} (as used by the MLE agent in Figure \ref{fig:length}).
Furthermore, the probability that adversaries observe 100 interaction histories before a forward secrecy transform is applied is ${\alpha^{100} = 10^{-100}}$.

\begin{figure}[t!]
    \centering
    \begin{subfigure}[t]{0.50\textwidth}
        \centering
        \includegraphics[
            width=\textwidth,
            trim={0cm, 0cm, 0cm, 0cm}, clip]{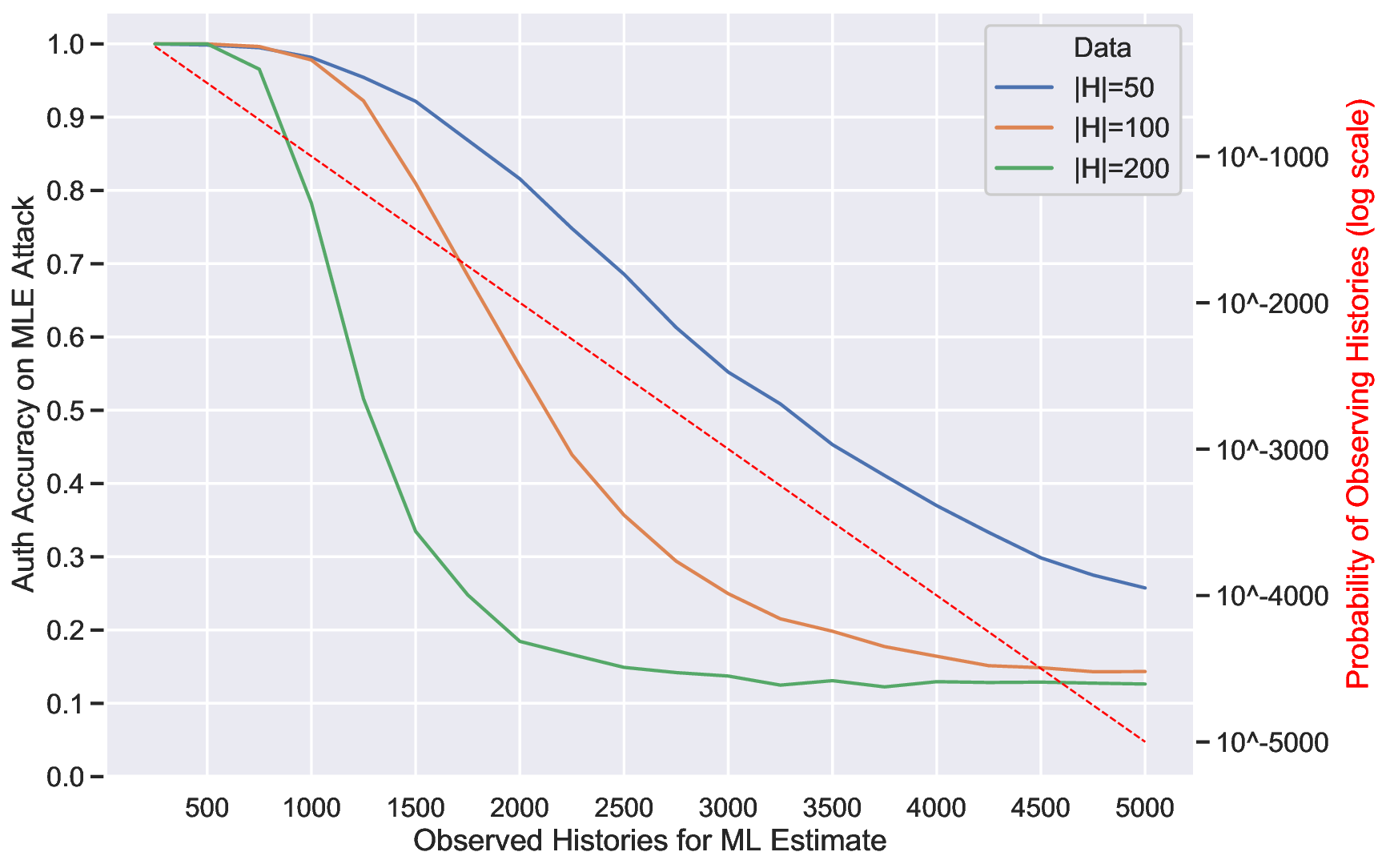}
        \caption{
        Effect of varying history length for fixed action space size $|\actionset|=10$.
        } 
        \label{fig:mle_rigor}
    \end{subfigure}
    \hfill 
    \begin{subfigure}[t]{0.45\textwidth}
        \centering
        \includegraphics[
            width=\textwidth,
            height=0.695\textwidth,
            scale=0.35
            ]
            {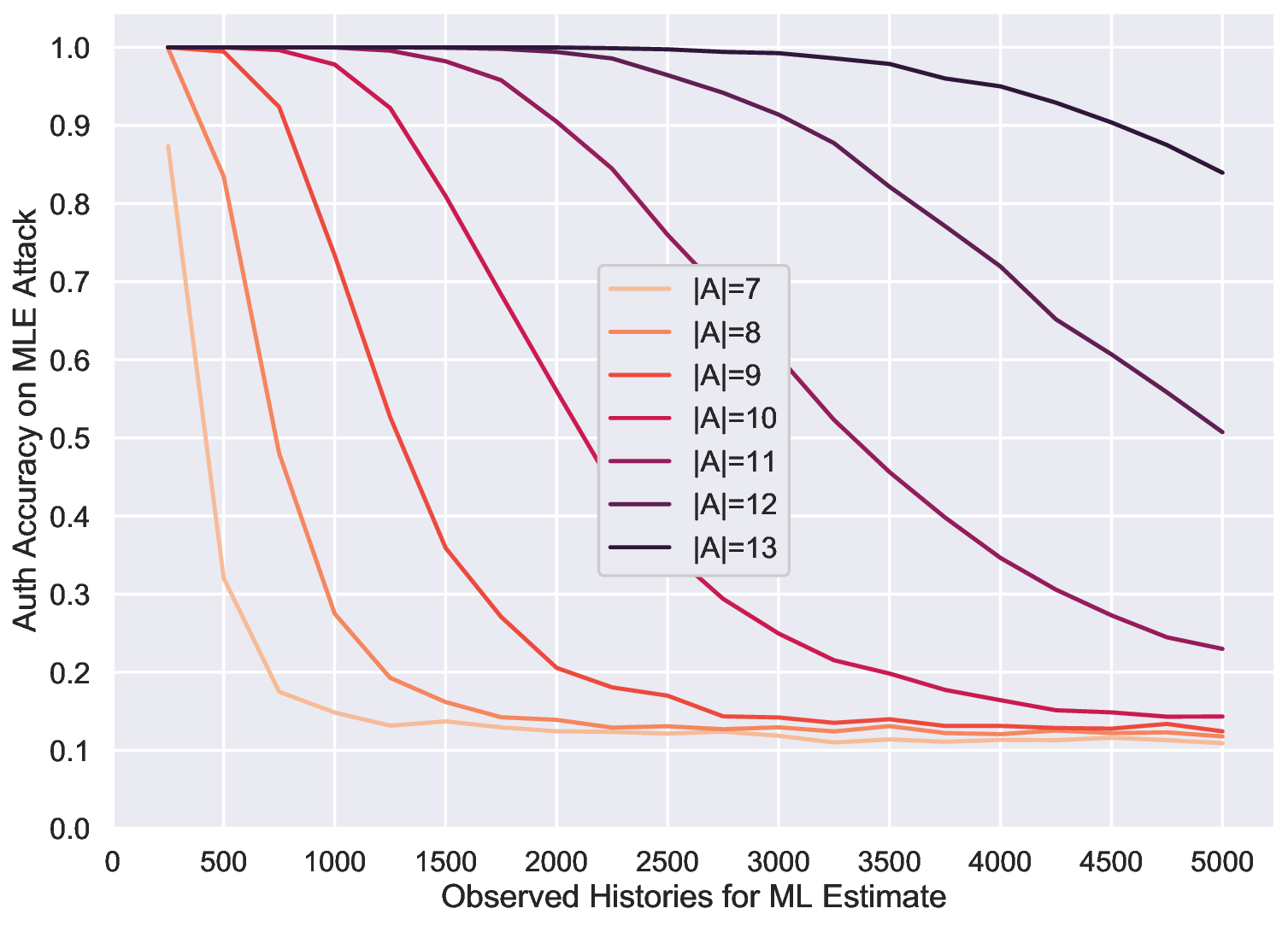}
        \caption{Effect of varying size of action space for fixed history length $|H_l| = 100$.
        }
        \label{fig:mle_vary_aspace}
    \end{subfigure}
    \caption{
    Average authentication accuracy against MLE attacks versus number of histories used for the MLE attack. Results averaged over 100 different client-server pairs. Standard error not shown due to low variation ($<0.01$).
    }
    \label{fig:mle_sec}
\end{figure}

We conduct an additional experiment to evaluate how many observed interactions are required for an MLE attack to obtain a high probability of authentication (with forward secrecy disabled).
Figure \ref{fig:mle_rigor} plots authentication accuracy on an MLE attack provided with an increasing number of histories. Results are averaged across 100 random client-server pairs, where accuracy is computed on 100 MLE histories for each pair. It also plots the probability of an intruder observing as many histories.
For longer history lengths ($|H_l|=200$), and with specified model complexity, \textit{at least 500 observed histories are required for an MLE attack to meaningfully lower the authentication accuracy}, and the probability of observing this much data before a forward secrecy transform is $10^{-500}$ under \ami. 
These results provide empirical evidence for the difficulty of constructing a successful attack from observed data, even by the best model estimation method, and without bounds on computational power. 

We also demonstrate that \ami's parameters can be tuned to further decrease the effectiveness of MLE attacks. In Figure \ref{fig:mle_vary_aspace}, we fix the history length at $|H_l|=100$, then vary the size of the action space $\actionset$ in the client and server PDT models.
The results show that larger action spaces -- corresponding to more complex models -- are more secure against MLE attacks in terms of number of histories the attacker must observe.

Finally, we include timing experiments for PyAMI's multi-agent interaction process. We measure the time for a full interaction history -- the transmission of all actions between separate machines -- to complete. For our experiments we use virtual machines on Google Compute Engine situated within the same geographic region (us-west1) and measure the time taken for a server-client interaction in the one-way authentication setup. 
For interaction lengths of $|H_l|=\{50,100,200\}$, we recorded interaction times of \{\SI{28(2)}{\milli\second}, \SI{54(2)}{\milli\second}, \SI{112(10)}{\milli\second}\} respectively, averaged over 100 trials.
These results show that \ami\ within PyAMI could be feasibly deployed to provide real-time authentication and key agreement.

\section{Optimizing Server Actions}
\label{optimize_server_actions}

Our empirical evaluation demonstrated that \ami\ robustly rejects various attack types while allowing legitimate clients to authenticate.
We now show how the server's decision model can be further optimized for protocol efficiency, as measured by the required interaction length before the $p$-value is sufficiently small to reject an adversary.
When the server interacts with an adversarial client, its actions can probe where the adversary may fail to match the legitimate client's action distributions.
Effective probing actions can lead to higher confidence decisions in shorter interaction history lengths.
Using shorter histories reduces the amount of observations adversaries can gather, thus improving the security of the protocol against model reconstruction attacks like the MLE attack.
We show how an effective server probing model can be learned for a given legitimate client model $\pi_u$ via reinforcement learning (\textsc{rl}) \cite{sutton1998reinforcement}.

We pose the server optimization problem as follows.
During training, the server decision model interacts with unknown clients over a series of length $l$ episodes in which each episode runs an \ami\ authentication process with a fixed client.
At the end of the interaction the server receives a reward, $R_l = 1 - p$ where $p$ is the $p$-value of the hypothesis test.
The server is rewarded for producing low $p$-values when interacting with adversaries.
The learning objective is thus:
\begin{equation}\label{eq:rl}
    \pi_s \in \arg\max_{\pi} \mathbf{E}_\pi \biggl [R_l \biggm | H_l \sim p_{s,c}, \pi_c \biggr],
\end{equation}
in which the client model, $\pi_c$, is sampled from an adversarial population (in our experiments we sample random \pdts\ the same way as Sec. \ref{sec:empirical:authentication}).
By applying an \textsc{rl} algorithm to optimize (\ref{eq:rl}) w.r.t. the server's decision model, we obtain a model that attempts to quickly reach high-confidence decisions.

We note that the server model is optimized with respect to a particular legitimate client model.
After successful authentication, the legitimate client model is transformed via a function $\phi$ so as to preserve forward secrecy.
In principle, this could render the server optimization obsolete since the client model has changed.
To address this concern, we can define $\phi$ to randomly permute the indexing of the client's actions at each leaf node of its \pdt. The random permutation generator is seeded by the session key, which depends on exact knowledge of the user model.
From an outside observer's perspective the distribution over elements of $\actionset$ will have changed, and is uniform on expectation assuming the permutation is sampled uniformly-randomly; thus an attacker could never learn anything but the uniform distribution over actions.
However, since the permutation is known to both the legitimate client and the server, the server model can un-permute the actions received from the client and apply the trained server model.

\section{Empirical Study: Optimized Probing}\label{sec:empirical:probing}

We conduct an empirical study to addresses the question: does effective probing lead to more efficient authentication relative to random probing?

In these experiments, the server model is a feedforward neural network which outputs the logits of a softmax distribution over the action space.
We use $|\mathcal{A}|=5$ and train with maximum interaction lengths of $50$ steps.
To more clearly show the benefit of server model optimization, we use legitimate client \pdt\ models with higher entropy action selection ($\tau=0.5$) than in Section \ref{sec:empirical:authentication}.
Such client models would be harder for an attacker to learn but also necessitate longer interaction histories for high confidence rejection decisions.
Thus, server policy optimization is more crucial to shorten the required interaction histories.

\begin{figure}[t]
     \centering
     \includegraphics[scale=0.35,width=0.5\textwidth]{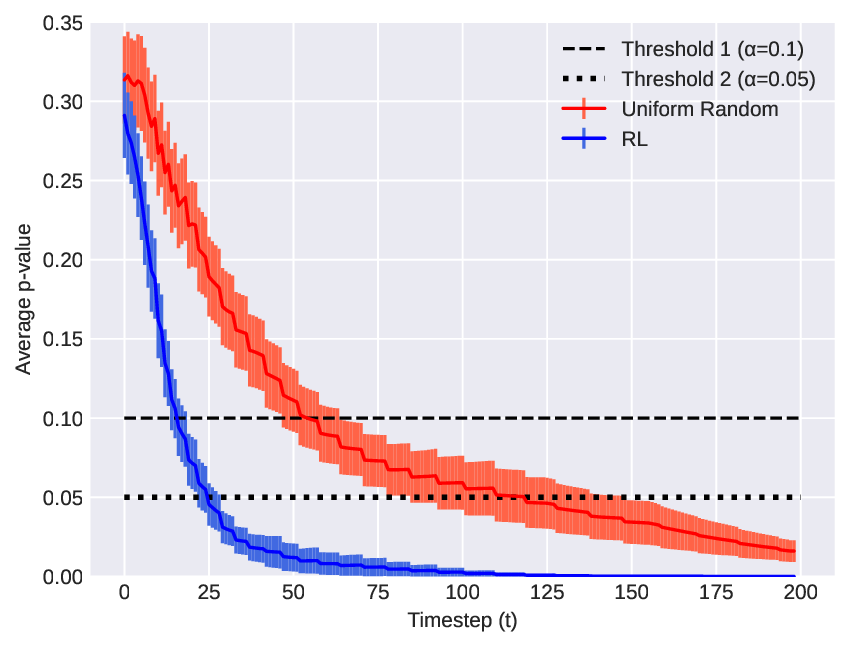}
     \caption{Average $p$-value per timestep over 10 different optimized servers interacting with Random adversaries.
     Shaded areas show standard error.\vspace{-10pt}}
     \label{fig:avg_pvalue_over_time}
\end{figure}

Using the \textsc{ppo} \textsc{rl} algorithm \cite{schulman2017proximal}, we train the server model for 5 million steps with 5,000 environment steps across three parallel processes for each model update.
We train the server for a fixed legitimate client model against an adversarial population of 100 randomly generated \pdts. 
After training, we evaluate the rate of $p$-value convergence for the trained server.
We compute the average $p$-value per timestep, averaged over a different population of 100 held-out adversarial \pdts.
As a baseline, we evaluate a uniform-random probing server model.
We repeat the server optimization process 10 times for different randomly generated legitimate clients (and unique populations of adversarial clients), to ensure our optimization method is effective not just for a specific server-client pair.

Figure \ref{fig:avg_pvalue_over_time} shows that the RL-trained server model leads to substantially faster convergence of p-values than uniform probing, reducing the required number of timesteps by 70\% and 79\% on average for  thresholds of $\alpha=0.1$ and $\alpha=0.05$, respectively.
The trained model is able to identify sequences of actions which lead to more informative observations for authenticating client agents.

\section{Conclusion and Future Work}
We contributed a novel protocol for secure authentication and key agreement based on abstract multi-agent interaction and agent modeling.
We have shown empirically that our protocol is highly accurate in authenticating legitimate users and rejecting different types of adversarial attacks. The protocol allows for control over authentication accuracy by choice of hypothesis test parameters, and by the chosen complexity of agent models. We released an open-source framework which employs our protocol in a distributed setting, and demonstrated the feasibility of this framework through timing experiments between remote server-client pairs. Finally, we showed how reinforcement learning can be used to train server models to achieve highly sample-efficient authentication.

Importantly, this work lays the ground work for multi-party authentication through multi-agent systems. Such a system raises new questions for how agents can jointly optimize security and efficiency; we believe that multi-agent reinforcement learning may offer a promising solution \cite{papoudakis2019dealing}. 
Future work could consider variable-length interaction histories, as such an authentication test could be more active in collecting additional information when facing decision uncertainty.

\bibliographystyle{splncs04}
\bibliography{security,post_quantum}

\end{document}